\documentclass[aps,prd,superscriptaddress,showpacs,tighten,nofootinbib]{revtex4}
\usepackage{epsfig}
\usepackage{amssymb}
\usepackage{rotate}
\usepackage{float}
\usepackage{graphicx}
\usepackage{slashed}
\usepackage{amsmath}
\usepackage{setspace}
\usepackage{slashed}
\usepackage{color}
\usepackage{comment}
\usepackage{dcolumn}   % needed for some tables
\usepackage{bm}        % for math
\usepackage{multirow}
\newcommand{\ba}{\begin{array}}
\newcommand{\ea}{\end{array}}
\newcommand{\bd}{\begin{displaymath}}
\newcommand{\ed}{\end{displaymath}}
\newcommand{\be}{\begin{equation}}
\newcommand{\ee}{\end{equation}}
\newcommand{\bea}{\begin{eqnarray}}
\newcommand{\eea}{\end{eqnarray}}

\def\th13 {\theta_{13}}

\usepackage{graphicx}% Include figure files

\newcommand{\mathsym}[1]{{}}
\newcommand{\unicode}[1]{{}}

\catcode`\@=11
\def\lsim{\mathrel{\mathpalette\@versim<}}
\def\gsim{\mathrel{\mathpalette\@versim>}}
\def\@versim#1#2{\vcenter{\offinterlineskip
\ialign{$\m@th#1\hfil##\hfil$\crcr#2\crcr\sim\crcr } }}
\catcode`\@=12

\catcode`@=12
\begin{document}

\title{Common Origin of Active and Sterile Neutrino Masses with Dark Matter}

\author{Rathin Adhikari}
\email{rathin@ctp-jamia.res.in}
\affiliation{Centre for Theoretical Physics, Jamia Millia Islamia (Central University), \\ Jamia Nagar, New Delhi-110025, India}

\author{Debasish Borah}
\email{dborah@tezu.ernet.in}
\affiliation{Department of Physics, Tezpur University,  \\ Tezpur - 784028, India}

\author{Ernest Ma}
\email{ma@physics.ucr.edu}
\affiliation{Department of Physics and Astronomy, University of California, \\  Riverside, CA 92521, USA}

\begin{abstract}
We propose an abelian extension of the Standard Model which can explain the origin of eV scale masses and mixing for active and sterile neutrinos and at the same time providing a natural cold dark matter candidate. One of the
three active neutrinos acquires mass at tree level through seesaw mechanism whereas the other two active neutrinos and one sterile
neutrino acquire eV scale masses at one-loop level. The model also allows non-trivial mixing between active and sterile neutrinos at one-loop level which could have interesting signatures at neutrino experiments. After the abelian gauge symmetry gets spontaneously broken down to a $Z_2$ symmetry, the lightest $Z_2$ odd particle can naturally give rise to the cold dark matter of the Universe. The phenomenology of both fermionic and scalar dark matter is briefly discussed by incorporating latest experimental constraints.
\end{abstract}
\maketitle
\newpage
\section{Introduction}
Light sterile neutrinos with mass at the electron Volt (eV) scale have gathered serious attention in the last few years after several experiments started suggesting additional light degrees of freedom beyond the three active neutrinos of the Standard Model. For a review, please see \cite{whitepaper}. The nine year Wilkinson Mass Anisotropy Probe (WMAP) data created lots of interest in light sterile neutrinos by constraining the number of light degrees of freedom to be$N_{\text{eff}} = 3.84 \pm 0.40$ \cite{wmap9}. This observation was however not fully supported by the recently reported data of the Planck collaboration which show a preference towards the standard three light neutrino picture with $ N_{\text{eff}} = 3.30^{+0.54}_{-0.51}$ \cite{planck}. Apart from the hints from cosmology experiments, there have also been evidence from anomalous results in accelerator and reactor based neutrino experiments. The anomalous results in anti-neutrino flux measurements at the LSND accelerator experiment \cite{LSND1} provided the first hint of light sterile neutrinos. The LSND results have also been supported by the latest data released by the MiniBooNE experiment \cite{miniboone}. Similar anomalies have also been observed at nuclear reactor neutrino experiments \cite{react} as well as gallium solar neutrino experiments \cite{gall}. These anomalies suggesting the presence of light sterile neutrinos have led to global short-baseline neutrino oscillation data favoring two light sterile neutrinos within the eV range \cite{Kopp:2011qd}. Some more interesting discussions on light sterile neutrinos from cosmology as well as neutrino experiments point of view can be found in \cite{cosmo_ste} and references therein. Thus, the hints in favor of sub-eV scale sterile neutrinos have led to a model building challenge to explain the origin of three light active neutrinos together with one or two sterile neutrinos within the same mass range. Some interesting proposals along these lines have appeared recently in \cite{rodejohann,model}. A nice review of some of the earlier works can also be found in \cite{merle}.

Recently, we discussed the possible origin of sub-eV scale active and sterile neutrino mass within the framework of a radiative neutrino mass model \cite{borahadhikari14}. In that work, an abelian gauge extention of the standard model, first discussed in \cite{Adhikari:2008uc} was suitably modified in order to accommodate eV scale sterile neutrino masses. The salient feature of the model is the way it relates dark matter with neutrino mass where neutrino masses arise at one loop level with dark matter particles running inside the loops: more popularly known as "scotogenic" model \cite{Ma:2006km}. The model allows the natural existence of a cold dark matter candidate which is stabilized by a remnant $Z_2$ symmetry after the abelian gauge symmetry is spontaneously broken. The dark matter phenomenology in this model was studied in \cite{Borah:2012qr} and more recently in \cite{boraharnab14} to explain the galactic center gamma ray excess. In the modified version of this model, as discussed in \cite{borahadhikari14}, it is possible to accommodate three active and one sterile neutrinos with eV scale masses. But to allow non-trivial mixing of light sterile neutrino with the active neutrinos, the remnamt $Z_2$ symmetry of the original model had to be sacrificed thereby losing the cold dark matter candidate. In the present work, we propose a modified version of this model which not only allows active and sterile neutrino masses and mixings at sub-eV scale but also keep the remnant $Z_2$ symmetry unbroken at low energy. Although our motivation in this work is to provide a common framework for sub-eV scale sterile neutrino and cold dark matter, it is straightforward to accommodate keV scale sterile neutrino as well which could give rise to scenario of warm dark matter.

This work is organized as follows. In section \ref{model}, we briefly discuss our model. In section \ref{sec:numass} and \ref{sec:stmass} we discuss the origin of sub-eV scale active and sterile neutrino masses respectively. We discuss the origin of active-sterile neutrino mixing in section \ref{sec:acstmass}. In section \ref{sec:darkmatter}, we briefly discuss the dark matter phenomenology in this model and then finally conclude in section \ref{sec:results}.
\section{The Model}
\label{model}
\begin{center}
\begin{table}
\begin{tabular}{|c|c|c|c|}
\hline
Particle & $SU(3)_c \times SU(2)_L \times U(1)_Y$ & $U(1)_X$ & $Z_2$ \\
\hline
$ (u,d)_L $ & $(3,2,\frac{1}{6})$ & $n_1$ & + \\
$ u_R $ & $(\bar{3},1,\frac{2}{3})$ & $\frac{1}{4}(7 n_1 -3 n_4)$ & + \\
$ d_R $ & $(\bar{3},1,-\frac{1}{3})$ & $\frac{1}{4} (n_1 +3 n_4)$ & +\\
$ (\nu, e)_L $ & $(1,2,-\frac{1}{2})$ & $n_4$ & + \\
$e_R$ & $(1,1,-1)$ & $\frac{1}{4} (-9 n_1 +5 n_4)$ & + \\
\hline
$N_R$ & $(1,1,0)$ & $\frac{3}{8}(3n_1+n_4)$ & - \\
$\Sigma_{1R,2R} $ & $(1,3,0)$ & $\frac{3}{8}(3n_1+n_4)$ & - \\
$ S_{1R}$ & $(1,1,0)$ & $\frac{1}{4}(3n_1+n_4)$ & + \\
$ S_{2R}$ & $(1,1,0)$ & $-\frac{5}{8}(3n_1+n_4)$ & + \\
\hline
$ (\phi^+,\phi^0)_1 $ & $(1,2,-\frac{1}{2})$ & $\frac{3}{4}(n_1-n_4)$ & + \\
$ (\phi^+,\phi^0)_2 $ & $(1,2,-\frac{1}{2})$& $\frac{1}{4}(9n_1-n_4)$ & + \\
$(\phi^+,\phi^0)_3 $ & $(1,2,-\frac{1}{2})$& $\frac{1}{8}(9n_1-5n_4)$ & - \\
\hline
$ \chi_1 $ & $(1,1,0)$ & $-\frac{1}{2}(3n_1+n_4)$ & + \\
$ \chi_2 $ & $(1,1,0)$ & $-\frac{1}{4}(3n_1+n_4)$ & - \\
$ \chi_3 $ & $(1,1,0)$ & $-\frac{3}{8}(3n_1+n_4)$ & - \\
$ \chi_4 $ & $(1,1,0)$ & $-\frac{3}{4}(3n_1+n_4)$ & + \\
$ \chi_5 $ & $(1,1,0)$ & $-\frac{1}{8}(3n_1+n_4)$ & + \\
\hline
\end{tabular}
\caption{Particle Content of the Model}
\label{table1}
\end{table}
\end{center}
To include both stable dark matter and sterile neutrinos along with active neutrinos our earlier proposed model \cite{borahadhikari14,Adhikari:2008uc} needs some modifications. The model which we consider has the following particle content as shown in table
\ref{table1}.  The third column in table \ref{table1} shows the $U(1)_X$ quantum numbers of various fields which satisfy the anomaly free conditions. The scalar fields content chosen above is not arbitrary and is needed, which leads to the possibility of radiative neutrino masses in a manner proposed in \cite{Ma:2006km} as well as a remnant $Z_2$ symmetry. Two more singlets $S_{1R}, S_{2R}$ are required to be present to satisfy the anomaly free conditions. In this model, the quarks couple to $\Phi_1$ and charged leptons to $\Phi_2$ whereas $(\nu, e)_L$ couples to $N_R, \Sigma_R$ through $\Phi_3$ and to $S_{1R}$ through $\Phi_1$. The four of extra five singlet scalars $\chi$ are needed to make sure that all the particles in the model acquire mass. The scalar singlet $\chi_5$ is required to include active and sterile neutrino mixing which
could explain LSND, MiniBooNe result. The lagrangian which can be constructed from the above particle content has an automatic $Z_2$ symmetry and hence provides a cold dark matter candidate in terms of the lightest odd particle under this $Z_2$ symmetry. Part of the scalar potential of this model relevant for our future discussion can be written as
$$ V_s \supset  \mu_2 \chi^2_2 \chi^{\dagger}_1 +\mu_3 \chi^2_3 \chi^{\dagger}_4 + \mu_4 \chi_1 \Phi^{\dagger}_1 \Phi_2 + \mu_5 \chi_3 \Phi^{\dagger}_3 \Phi_2 +\lambda_{13} (\Phi^{\dagger}_1 \Phi_1)(\Phi^{\dagger}_3 \Phi_3)$$
$$  +f_1 \chi_2 \chi^{\dagger}_5\Phi^{\dagger}_1\Phi_3+f_3 \chi_1 \chi^{\dagger}_3\Phi^{\dagger}_1\Phi_3 +f_4 \chi^2_2\Phi^{\dagger}_1\Phi_2 + f_5 \chi^{\dagger}_3\chi_4 \Phi^{\dagger}_3 \Phi_2 $$
\begin{equation}
 +\lambda_{23} (\Phi^{\dagger}_2 \Phi_2)(\Phi^{\dagger}_3 \Phi_3) + \lambda_{15} (\Phi^{\dagger}_1 \Phi_1)(\chi^{\dagger}_2 \chi_2)
+ \lambda_{16} (\Phi^{\dagger}_1 \Phi_1)(\chi^{\dagger}_3 \chi_3) + \lambda_{26} (\Phi^{\dagger}_2 \Phi_2)(\chi^{\dagger}_3 \chi_3)
\label{scalpot}
\end{equation}
The vacuum expectation values (vev) of various scalar fields are denoted as $ \langle \phi^0_{1,2} \rangle = v_{1,2}, \; \langle \chi^0_{1,4,5} \rangle  =u_{1,4,5}$ and the coupling constants of $SU(2)_L, U(1)_Y, U(1)_X$ are denoted as $g_2, g_1, g_x$ respectively. The mass of charged weak bosons is given by $M^2_W = \frac{g^2_2}{2}(v^2_1+v^2_2) $. In the $(W^{\mu}_3, Y^{\mu}, X^{\mu})$ basis the 
neutral gauge boson mass matrix is given by
\begin{equation}
M =\frac{1}{2}
\left(\begin{array}{cccc}
\ g^2_2(v^2_1+v^2_2) & g_1g_2(v^2_1+v^2_2) &  M^2_{WX} \\
\ g_1g_2(v^2_1+v^2_2) & g^2_1(v^2_1+v^2_2) & M^2_{YX} \\
\ M^2_{WX} & M^2_{YX}  & M^2_{XX}
\end{array}\right)
\end{equation}
where 
$$M^2_{WX} = -g_2g_x(\frac{3}{4}(n_1-n_4)v^2_1+\frac{1}{4}(9n_1-n_4)v^2_2) $$
$$ M^2_{YX} = -g_1g_x(\frac{3}{4}(n_1-n_4)v^2_1+\frac{1}{4}(9n_1-n_4)v^2_2)$$
$$ M^2_{XX} = g^2_x(\frac{9}{4}(n_1-n_4)^2v^2_1+\frac{1}{4}(9n_1-n_4)^2v^2_2+\frac{1}{64}(3n_1+n_4)^2(16u^2_1+36u^2_4+u^2_5)) $$
From the above mass matrix one can see that there is  mixing between the electroweak gauge bosons and the additional $U(1)_X$ boson. But it should be very small as required by the electroweak precision measurements. To avoid stringent constraint on mixing one may assume a very simplified framework where there is no mixing between the electroweak gauge bosons and the extra $U(1)_X$ boson. This implies $ M^2_{WX} = M^2_{YX} = 0$ resulting in the following relation:
\begin{equation}
3(n_4-n_1)v^2_1 = (9n_1-n_4)v^2_2 
\label{zeromixeq}
\end{equation}
which can be satisfied if $1 < n_4/n_1 <9 $. As discussed in  \cite{Adhikari:2008uc} if $U(1)_X$ boson is observed at LHC, the ratio $n_4/n_1$ could be found empirically 
from its' decay to $q\bar{q}$, $l\bar{l}$ and $\nu\bar{\nu}$ where $q$, $l$ and $\nu$ correspond to 
quarks, charged leptons and neutrinos respectively.
The vevs $v_1$ and $v_2$ can be written in terms of the charged weak boson mass as 
$$ v^2_1 = \frac{M^2_W(9n_1-n_4)}{g^2_2(3n_1+n_4)}, \quad v^2_2 = \frac{M^2_W(-3n_1+3n_4)}{g^2_2(3n_1+n_4)} $$
Assuming zero mixing, the neutral gauge bosons of the Standard Model- the photon and weak Z boson have masses
$$ M_B = 0, \quad M^2_Z = \frac{(g^2_1+g^2_2)M^2_W}{g^2_2} $$
respectively. The mass of $U(1)_X$ gauge boson mass is given by
$$ M^2_X = 2g^2_X (-\frac{3M^2_W}{8g^2_2}(9n_1-n_4)(n_1-n_4)+\frac{1}{64}(3n_1+n_4)^2(16u^2_1+36u^2_4+u^2_5)) $$
For simplicity, we assume $u_1 = u_4 = u_5 = u$ such that the mass of $X$ boson can be written as
\begin{align}
M^2_X &= 2g^2_X\bigg{[}-3\frac{m^2_W}{8g^2_2}(9n_1-n_4)(n_1-n_4)+\frac{53}{64}(3n_1+n_4)^2u^2\bigg{]}
\end{align}

\section{Active Neutrino Mass}
\label{sec:numass}

Here we discuss the origin of three active neutrino mass in the model. The relevant part of the Yukawa Lagrangian is 
$$ \mathcal{L}_Y \supset y \bar{L} \Phi^{\dagger}_1 S_{1R} + h_N \bar{L} \Phi^{\dagger}_3 N_R + h_{\Sigma}  \bar{L}\Phi^{\dagger}_3 \Sigma_R + f_N N_R N_R \chi_4+ f_S S_{1R} S_{1R} \chi_1 $$
\begin{equation}
+ f_{\Sigma} \Sigma_R \Sigma_R \chi_4 + f_{NS} N_R S_{2R} \chi^{\dagger}_2 
\label{yukawa} 
\end{equation}

\begin{figure}[htb]
\centering
\includegraphics[scale=0.75]{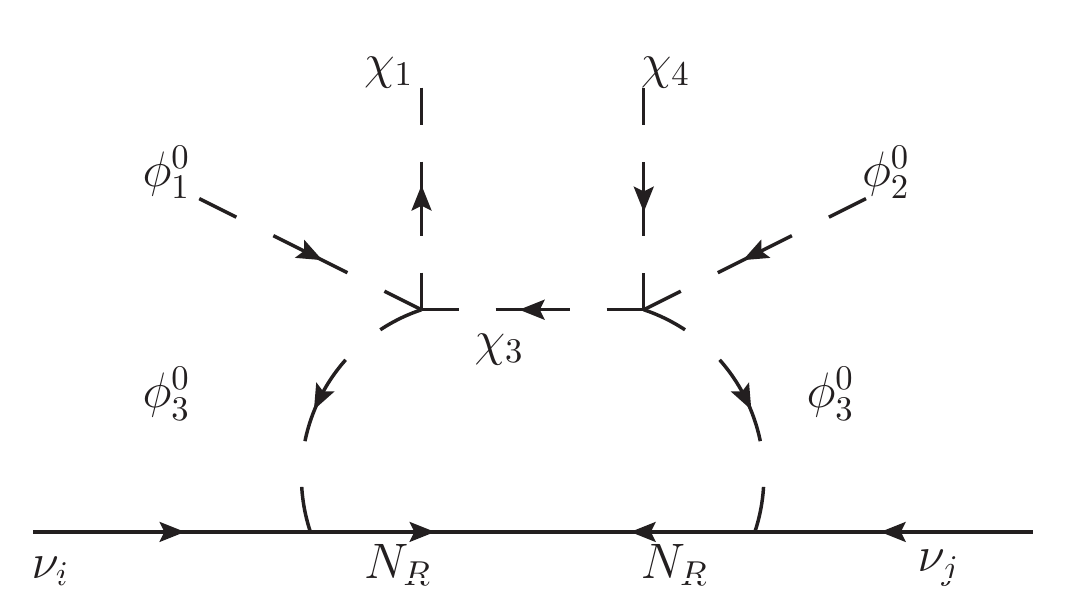}
\caption{One-loop contribution to active neutrino mass}
\label{numass}
\end{figure}

The fields $\chi_{1,2,4}$ has non-zero vev and there is spontaneous symmetry breaking of $U(1)_X$ symmetry. This results in the Majorana mass of the fermions $S_R, N_R$ and $\Sigma_R$. Active neutrinos  couples to $S_{1R}$ as shown in equation $(\ref{yukawa})$. The 
$S_{1R}$  fermion acquire heavy mass due to vev of $\chi_{1}$. Thus the $3 \times 3$ active neutrino mass matrix receives tree level contribution from standard type I seesaw mechanism.   $S_{1R}$ can couple to arbitrary linear combination of $\nu_i$ by assigning different values of $y$ in
equation (\ref{yukawa}) for
different generation. However,  if we consider this coupling non-zero  and same for $\nu_\mu$ and $\nu_\tau$ only then  the heaviest mass $m_{\nu 3}$ is 
given by 
\begin{equation}
m_{\nu 3} \approx \frac{ 2y^2 v_1^2}{f_S u_1}
\label{neutmass}
\end{equation}
and one gets the hierarchical pattern of neutrino mass with only one massive and other two neutrinos massless at the tree level. This sets the scale of higher neutrino mass square difference of  about $ 2.4  \times 10^{-3}$ eV$^2$ and for that
$m_{\nu 3}  $ may be considered to be about $0.05$ eV (for hierarchical neutrino masses) to about 0.1 eV (for almost degenerate neutrino masses).
Two neutrinos which are massless at the tree level become massive from one loop contribution 
as shown in figure \ref{numass} of Feynman diagram in which  any one of one singlet $N_R$ and two triplet $\Sigma^0_{1R, 2R}$ couples to
active neutrinos. These one loop contributions 
set the scale of lower mass squared difference of about $7.6 \times 10^{-5}$ eV$^2$ for which other neutrino masses may be considered to be about $10^{-2}$ eV (for hierarchical neutrino masses ) to about 0.1 eV (for almost degenerate neutrino masses). Somewhat similar to \cite{Ma:2006km}, such a one loop diagram gives partial
contribution through $A_k$ as mentioned below when there is a mass splitting between the CP-even and CP-odd neutral components of the Higgs field involved in the loop which is $\phi^0_3$ in this case. Such a mass splitting is possible due to the  couplings between $\phi^0_3$ and the singlet scalar fields $\chi_i$ shown in equation $(\ref{scalpot})$ and is necessary for $\phi^0_3$ to be a dark matter candidate as discussed in \cite{Borah:2012qr}.

The  one-loop contribution $(M_\nu)_{ij}$ to $3\times 3$ active neutrino mass matrix is obtained as

\begin{eqnarray} 
({M_\nu)}_{ij} \approx  \frac{f_3 f_5 v_1 v_2 u_1 u_4}{16 \pi^2} \sum_k {h_{N, \Sigma} }_{ik} {h_{N, \Sigma} }_{jk} \left( A_k +{(B_k)}_{ij} \right)
\label{nuradmass}
\end{eqnarray}
where $k=1,2,3$ corresponds to different possible fermions  $N_R$, $\Sigma_{1R}$ and $\Sigma_{2R}$ respectively which could be present in
one loop diagram. Here, $A_k$ and $B_k$ are given by 
\begin{eqnarray}
A_k &=& {(M_{N, \Sigma})}_k \left[ I\left( m_{\phi^0_{3R}},m_{\phi^0_{3R}},{(M_{N, \Sigma})}_k, m_{\chi_{3R}} \right) - I\left(m_{\phi^0_{3I}},m_{\phi^0_{3I}},{(M_{N, \Sigma})}_k, m_{\chi_{3R}} \right) \right], 
\label{Ak}
\end{eqnarray}
\begin{eqnarray}
{(B_k)}_{ij}= -(2-\delta_{ij}) {(M_{N, \Sigma})}_k I\left(m_{\phi^0_{3R}},m_{\phi^0_{3I}},{(M_{N, \Sigma})}_k, m_{\chi_{3I}} \right),
\label{Bk}
\end{eqnarray}
in which
\begin{eqnarray}
I(a,a,b,c)=  \frac{(a^4-b^2 c^2) \ln (a^2/c^2)}{{(b^2 -a^2)}^2 {(c^2 -a^2)}^2}+ \frac{b^2 \ln (b^2/c^2)}{ (c^2 - b^2){(a^2 - b^2)}^2}  
  -\frac{1}{(a^2 - b^2) (a^2 -c^2)},
\end{eqnarray}
\begin{eqnarray} 
I(a,b,c,d)&=&  \frac{1}{a^2-b^2}  \left[  \frac{1}{a^2-c^2} \left( \frac{a^2}{a^2-d^2} \ln(a^2/d^2) - \frac{c^2}{c^2-d^2} \ln(c^2/d^2)\right)\right. \nonumber \\  &-&  \left. \frac{1}{b^2-c^2} \left( \frac{b^2}{b^2-d^2} \ln(b^2/d^2) - \frac{c^2}{c^2-d^2} \ln(c^2/d^2)\right)  \right]
\end{eqnarray}
and
$m_{\phi^0_{3R}}$ and $m_{\phi^0_{3I}} $ are the masses corresponding to $Re[\phi^0_3]$ and $Im[\phi^0_3]$ respectively
and  $m_{\chi_{3R}}$ and  $m_{\chi_{3I}}$ are the masses corresponding to $Re[\chi^0_3]$ and $Im[\chi^0_3]$ respectively  and 
$v_i= \langle \phi_i \rangle$ and $u_i= \langle \chi_i \rangle$ . 
$M_{N,\Sigma}$ is the Majorana mass term of $N_R(\Sigma^0_R)$. $h_{N, \Sigma}$ are the Yukawa couplings in equation $(\ref{yukawa})$. 
If $N_R$ is replaced by $\Sigma_R$ in the one loop diagram then in the above expression $h_{N_{ij}}$ is replaced by $h_{\Sigma_{ij}}$.

One may write $A_k$ and $B_k$ in somewhat simpler forms as shown below. We
write  $(M_{N, \Sigma})_k$  as $m_{2k}$.
Ignoring the mixing between $\phi_3^0$ and $\chi_3^0$,
and considering all the scalar masses in the loop diagram almost degenerate and written as $m_{sc}$ one obtains 
\begin{eqnarray}
A_k + (B_k)_{ij} \approx m_{2k} \left[\frac{m_{sc}^2 
+ m_{2k}^2 }{m_{sc}^2 \left( m_{sc}^2 - m_{2k}^2 \right)^2 }- \frac{(2-\delta_{ij})\; m_{2k}^2}{\left(m_{sc}^2 - m_{2k}^2 \right)^3}\ln \left( m_{sc}^2/m_{2k}^2 \right)   \right],
\label{scaldeg}
\end{eqnarray}
and if all scalar and fermion masses in the loop are almost degenerate and written as $m_{deg}$ then
\begin{eqnarray}
A_k + (B_k)_{ij} \approx  \frac{(2-\delta_{ij})}{6 m_{deg}^3}\; .
\label{fermscal}
\end{eqnarray}
With appropriate choices of the couplings and vevs one can get suitable neutrino masses (as mentioned earlier for hierarchical or almost 
degenerate neutrinos). 

\section{Sterile Neutrino Mass}
\label{sec:stmass}
\begin{figure}[h!]
\centering
%\begin{tabular}{cc}
\includegraphics[width=0.5\textwidth]{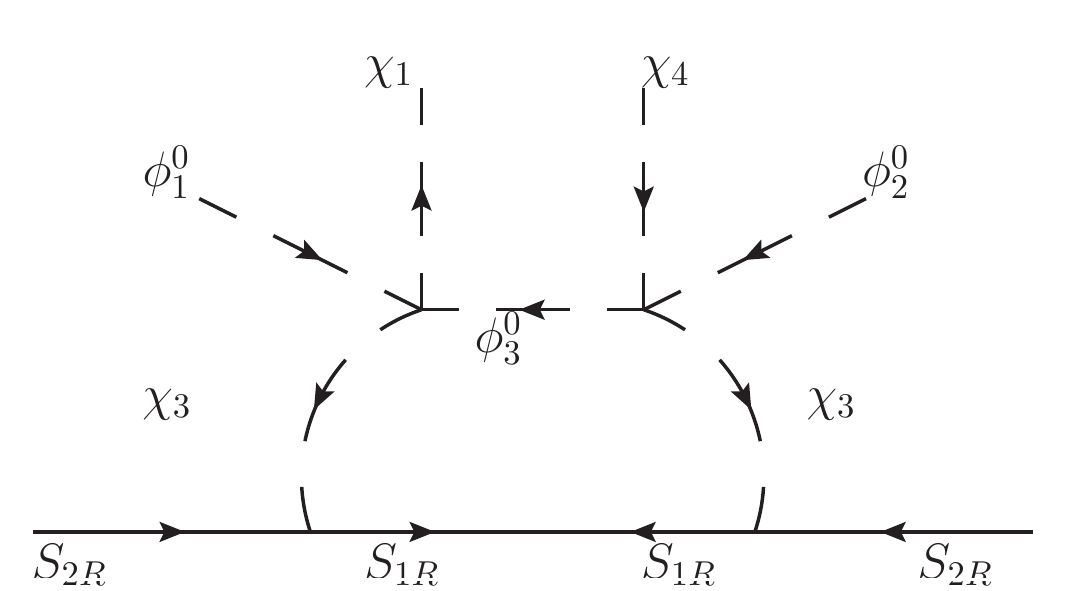}
%\includegraphics[width=0.5\textwidth]{activesterile3}
%\end{tabular}
\caption{One-loop contribution to sterile neutrino mass}
\label{sterile1}
\end{figure}
\begin{figure}[h!]
\centering
%\begin{tabular}{cc}
\includegraphics[width=0.5\textwidth]{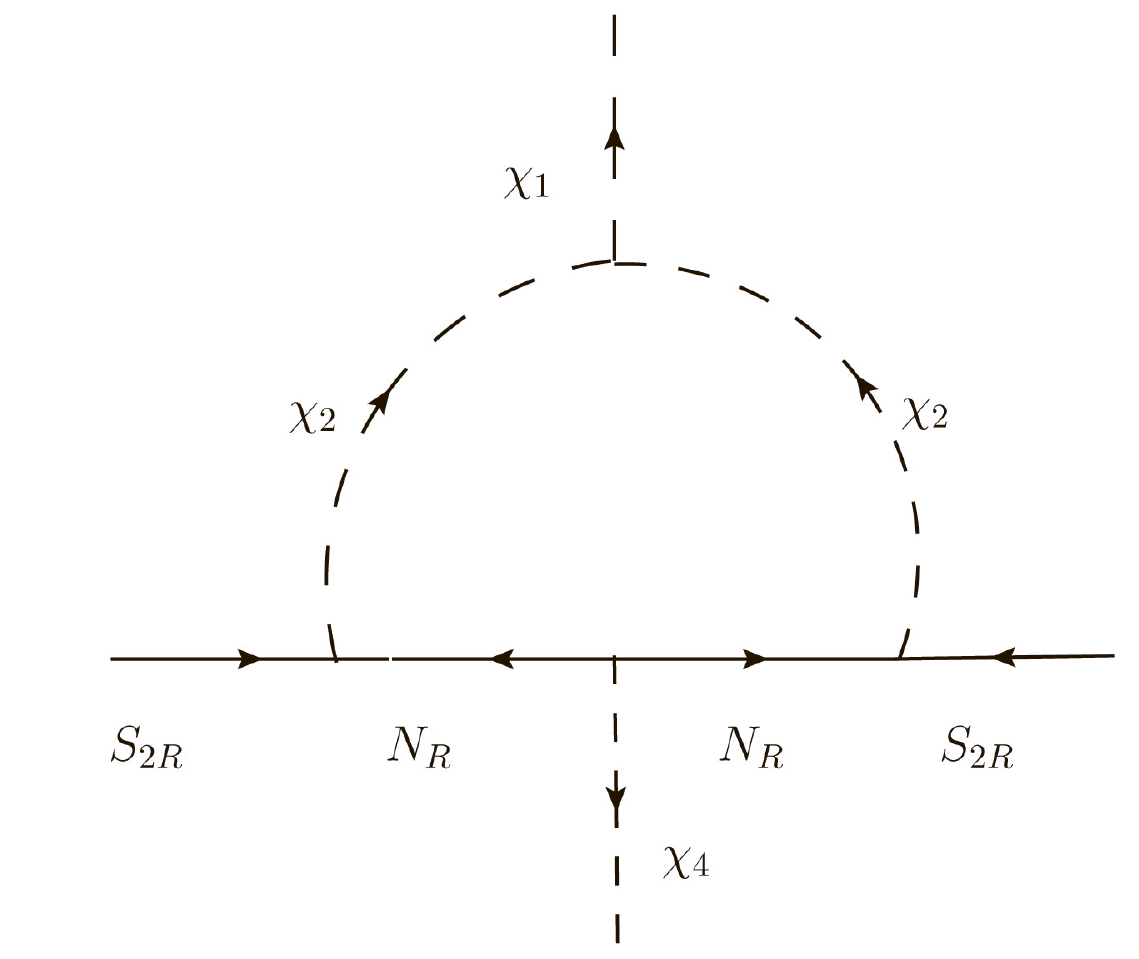}
%\includegraphics[width=0.5\textwidth]{activesterile3}
%\end{tabular}
\caption{One-loop contribution to sterile neutrino mass}
\label{sterile2}
\end{figure}

There are three singlet fermions $S_{1R}, S_{2R}, N_R$ in the model. Out of these, $S_{1R}$ couples to the active neutrinos and contribute to tree level mass term of one of the active neutrinos through seesaw mechanism and for generic Dirac Yukawa couplings of the neutrinos, the sterile neutrino $S_{1R}$ is expected to be much heavier than the eV scale.  Thus either $S_{2R}$ or $N_R$ or both could give rise to the light sterile neutrinos. From the Yukawa lagrangian \ref{yukawa}, one can see that there is a tree level mass term $f_N \langle \chi_4 \rangle$  for $N_R$ which is
 much above eV order as we are considering the additional $U(1)_X$ symmetry to be broken (by the vev of $\chi$) at a scale higher than the electroweak scale. However, $S_{2R}$ can be a good candidate for sterile neutrino as there is no tree level mass term for it. There are mixing terms of $S_{2R}$ with $N_R, S_{1R}$ through scalar fields $\chi_2, \chi_3$ respectively. However, the fields $\chi_2$ and $\chi_3$ does not acquire vev as they are $Z_2$-odd and at the tree level $S_{2R}$ decouples from rest of the fermions. At the tree level
 the full neutrino mass matrix is written as
\begin{equation}
M_f =
\left(\begin{array}{cccccc}
\ 0 & 0 & 0 & y_1v_1 & 0 & 0 \\
\ 0 & 0 & 0 & y_2v_1 & 0 & 0 \\
\ 0 & 0 & 0 & y_3 v_1 & 0 & 0 \\
\ y_1v_1 & y_2v_1 & y_3v_1 & f_S u_1 & 0 & 0 \\
\ 0 & 0 & 0 & 0 & 0 & 0 \\
\ 0 & 0 & 0 & 0 & 0 & f_N u_4
\end{array}\right)
\label{numassmatrix}
\end{equation}
in the basis $(\nu_e, \nu_{\mu}, \nu_{\tau}, S_{1R}, S_{2R}, N_R)$. The sterile neutrino $S_{2R}$ which is massless at tree level,
can aquire a small eV scale mass at one loop level from the diagrams \ref{sterile1} and \ref{sterile2}. The mass term for  $S_{2R}$ corresponds to  $(M_f)_{55}$ element in the above basis. The one-loop contribution to this element is obtained as
\begin{eqnarray} 
{(M_f)}_{55} \approx  \frac{f_{12}^2 f_3 f_5 v_1 v_2 u_1 u_4}{16 \pi^2}  \left( A +B \right) \nonumber \\
+ \mu_2 {f_{NS}}^2 u_1 M_X I(m_{\chi_2},M_X)
\label{eq:nuradmass1}
\end{eqnarray}
where $A$ and $B$ can be obtained by replacing ${(M_{N, \Sigma})}_k$ by $M_{S_{1R}}$ in $A_k$ and $B_k$ in $(\ref{Ak})$ and $(\ref{Bk})$ respectively and
\begin{eqnarray} 
I\left( a,b\right) =  \frac{1}{{(a^2-b^2)}^2}\left( a^2-b^2 - b^2 \ln (a^2/b^2) \right)\; .
\end{eqnarray}
With suitable choices of the couplings and the vevs it is possible to get ev scale sterile neutrino mass like other active neutrino masses.

\section{Active Sterile Neutrino Mixing}
\label{sec:acstmass}
\begin{figure}[htb]
\centering
\includegraphics[width=0.5\textwidth]{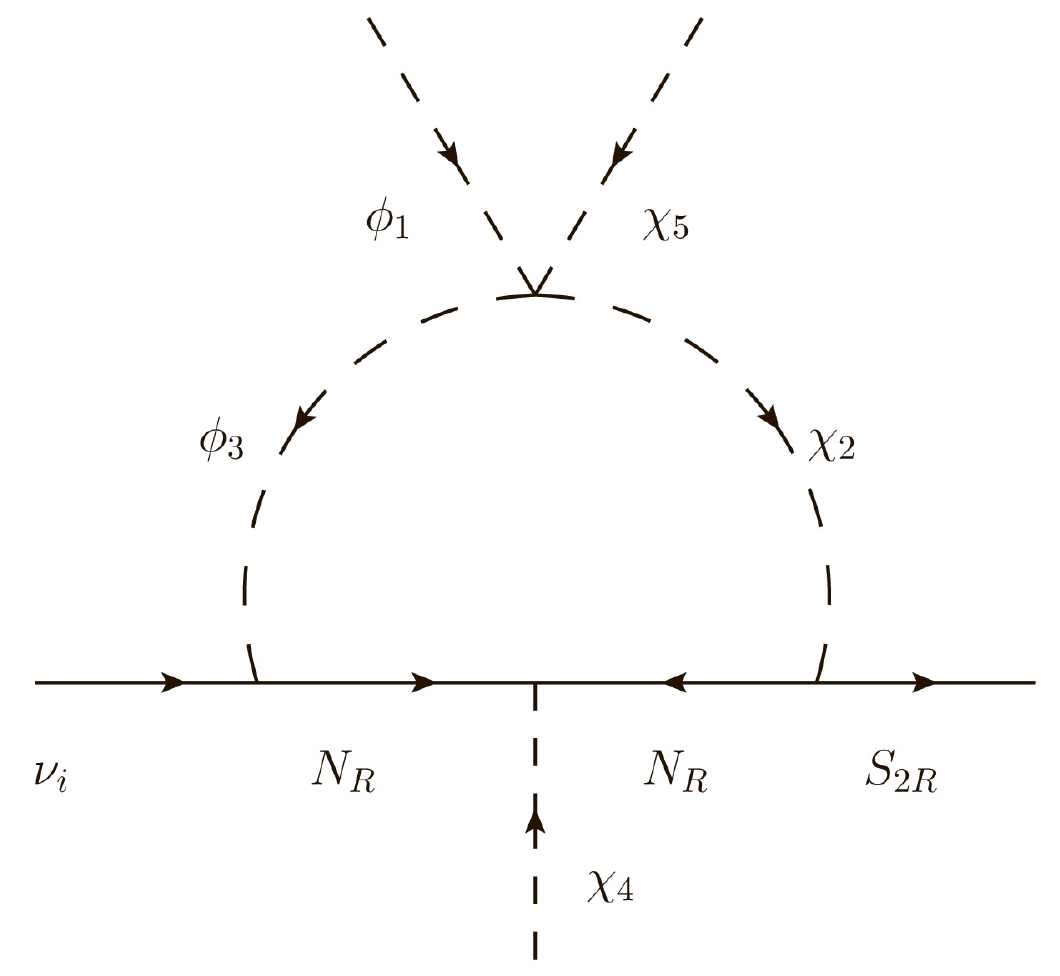}
\caption{One-loop contribution to active-sterile neutrino mixing}
\label{actsterile1}
\end{figure}

Although the sterile neutrinos could have light mass but  to have implications in neutrino oscillation experiments, there should be
some non-trivial mixing of sterile neutrino with the active neutrinos. There is no tree level mixing of  $S_{2R}$  with the active neutrinos. But at one loop level, there is  active sterile neutrino mixing as shown in figure \ref{actsterile1}. 
 The mixing term  can be estimated as
\begin{eqnarray} 
{(M_f)}_{j5} = {(M_f)}_{5j}^* \approx   \frac{f_{1} f_{NS}  v_1 u_5 (h_{N})_{j} M_X }{16 \pi^2} \left[  I\left(m_{\chi_{2R}},m_{\phi_{3R}^0}, M_X\right)-I\left(m_{\chi_{2I}},m_{\phi_{3I}^0}, M_X\right)\right]  \nonumber \\
\label{nuradmass2}
\end{eqnarray}
where $j = 1,2,3$ and
$M_X$ is the $U(1)_X$ symmetry breaking scale and 
\begin{eqnarray} 
I\left( a,b,c\right) =  \frac{a^2 b^2 \ln(a^2/b^2) + b^2 c^2 \ln(b^2/c^2)+c^2 a^2 \ln(c^2/a^2)}{(a^2-b^2) (b^2 -c^2) (c^2-a^2)}\; .
\end{eqnarray}
Here, we have assumed 
$  M_{N_R} \sim M_X $. If we assume   $m_{\chi_{2R}} \sim m_{\phi_{3R}^0}$ and/or $m_{\chi_{2I}} \sim m_{\phi_{3I}^0}$ then to get ${(M_f)}_{5j} $ from \eqref{nuradmass2}
one is required to consider $I\left( a,b\right)$ mentioned earlier instead of $I\left( a, b,c \right) $ in \eqref{nuradmass2}.  
If we assume $m_{\chi_{2R}} \sim m_{\chi_{2I}} \ll M_X$, then 
\begin{equation} 
{(M_f)}_{j5}\approx    \frac{f_{1} f_{NS}  v_1 u_5 (h_{N})_{j} M_X }{16 \pi^2 M^4_X} \left[ m^2_{\chi_{2R}}-m^2_{\chi_{2I}}-M^2_X \ln ( m^2_{\chi_{2R}}/m^2_{\chi_{2I}})\right]\; .
\end{equation}
Writing $m_{\chi_2}= (m_{\chi_{2R}}+m_{\chi_{2I}})/2$  
\begin{equation}
{(M_f)}_{15}\approx  \frac{f_{1} f_{NS}  v_1 u_5 (h_{N})_{1} }{16 \pi^2 M_X m_{\chi_2}^2 } \left[ m^2_{\chi_{2R}}-m^2_{\chi_{2I}}\right]
\end{equation} 
For ${(M_f)}_{25}$ in the above expression $(h_{N})_{1}$ will be replaced by $(h_{N})_{2}$.  As 
$A+B \approx 1/( m_{\chi_3}^2 M_X)$ 
\begin{eqnarray} 
{(M_f)}_{55} \approx  \frac{f_{12}^2 f_3 f_5 v_1 v_2 u_1 u_4}{16 \pi^2 m_{\chi_3}^2 M_X } + \frac{\mu_2 f_{NS}^2 u_1 }{16 \pi^2 M_X
m_{\chi_2}^2} \left[ m^2_{\chi_{2R}}-m^2_{\chi_{2I}}\right]
\end{eqnarray}
The first term and 2nd term on right hand side of above equation corresponds to 1st and 2nd term of equation \eqref{eq:nuradmass1}.
Which of these two terms will dominate, depends on the details of the model. We shall assume the first term to dominate 
in the later part of discussion.
As  ${(M_f)}_{11} \sim {(M_f)}_{22} \ll {(M_f)}_{55}$ we can write the active sterile mixing angles $\theta_{e5}$  and $\theta_{\mu 5}$ as 
\begin{equation}
\tan 2\theta_{e5} = \frac{2{(M_f)}_{15}}{{(M_f)}_{55}} ;\;\;\; \tan 2\theta_{\mu 5} = \frac{2{(M_f)}_{25}}{{(M_f)}_{55}} 
\end{equation}
 One may note that unlike the conventional 4th row and  column for the sterile entry in the neutrino mass matrix we have considered  it in the
5th entry. As $\theta_{e5}$ and $\theta_{\mu 5}$ are very small (following the conventional notation of 4th entry for sterile neutrino in writing
neutrino mixing matrix elements), if we assume $u_4 \sim u_5$ then  
\begin{equation}
|U_{e4}| \sim  \sin\theta_{e5} \approx \frac{1}{2} \tan 2\theta_{e5} = \frac{ f_1 f_{NS} (h_{N})_{1} m_{\chi_3}^2 }{f_{12}^2 f_3 f_5 v_2 u_1 m_{\chi_2}^2} \left[ m^2_{\chi_{2R}}-m^2_{\chi_{2I}}\right] \; .
\label{sterilemix}
\end{equation}
$|U_{\mu 4}|$ is like the above  expression with the replacement of  $(h_{N})_{1}$  by $(h_{N})_{2}$. 
The global neutrino fit data with three active and one light sterile neutrinos \cite{sterileglobal} give the best fit parameters as $\Delta m^2_{41} = 0.93 \; \text{eV}^2, \; \lvert U_{e4} \rvert = 0.15, \; \lvert U_{\mu 4} \rvert  = 0.17$. One can satisfy these requirements easily within the framework of our model. As for example, considering all Yukawa couplings of the order of $10^{-3}$ and mass splitting between the real and imaginary component of the fields $\chi_3, \phi^0_3$ such that $(m^2_{\chi_{3R}}-m^2_{\chi_{3I}}) \approx 100 \; \text{GeV}^2 $ and the $vev$'s $v_2 \approx u_1 \approx 200 \; \text{GeV}$ and $m_{\chi_3}^2 \sim 0.1 \times m_{\chi_2}^2$  one may 
obtain the required order for active sterile mixing.  Using the experimental data the two couplings $(h_{N })_{1}$ and $(h_{N})_{2}$ 
can be related as follows:
\begin{equation}
\frac{(h_{N })_{1}}{(h_{N})_{2}} \approx  \frac{|U_{e4}| }{|U_{\mu 4}| }  \approx \frac{0.15}{0.17}
\end{equation}
and they are almost equal.

Is there scope to consider the sterile neutrino as dark matter candidate ? Due to the radiative origin of the mass of $S_{2R}$ as given by equation \eqref{eq:nuradmass1} with appropriate choice 
of vevs and the couplings its' mass could be of the order of keV. But it is not odd under $Z_2$ and as such not protected 
by the symmetry to be stable. But as it could be much lighter than $N_R$ as well as $\chi_2$ so due to the kinemetic  constraint it could be
stable and interestingly could play the role of dark matter with relatively smaller active sterile mixing. 

\section{Dark Matter}
\label{sec:darkmatter}
\begin{figure}[htb]
\centering
\begin{tabular}{cc}
\includegraphics[width=0.5\textwidth]{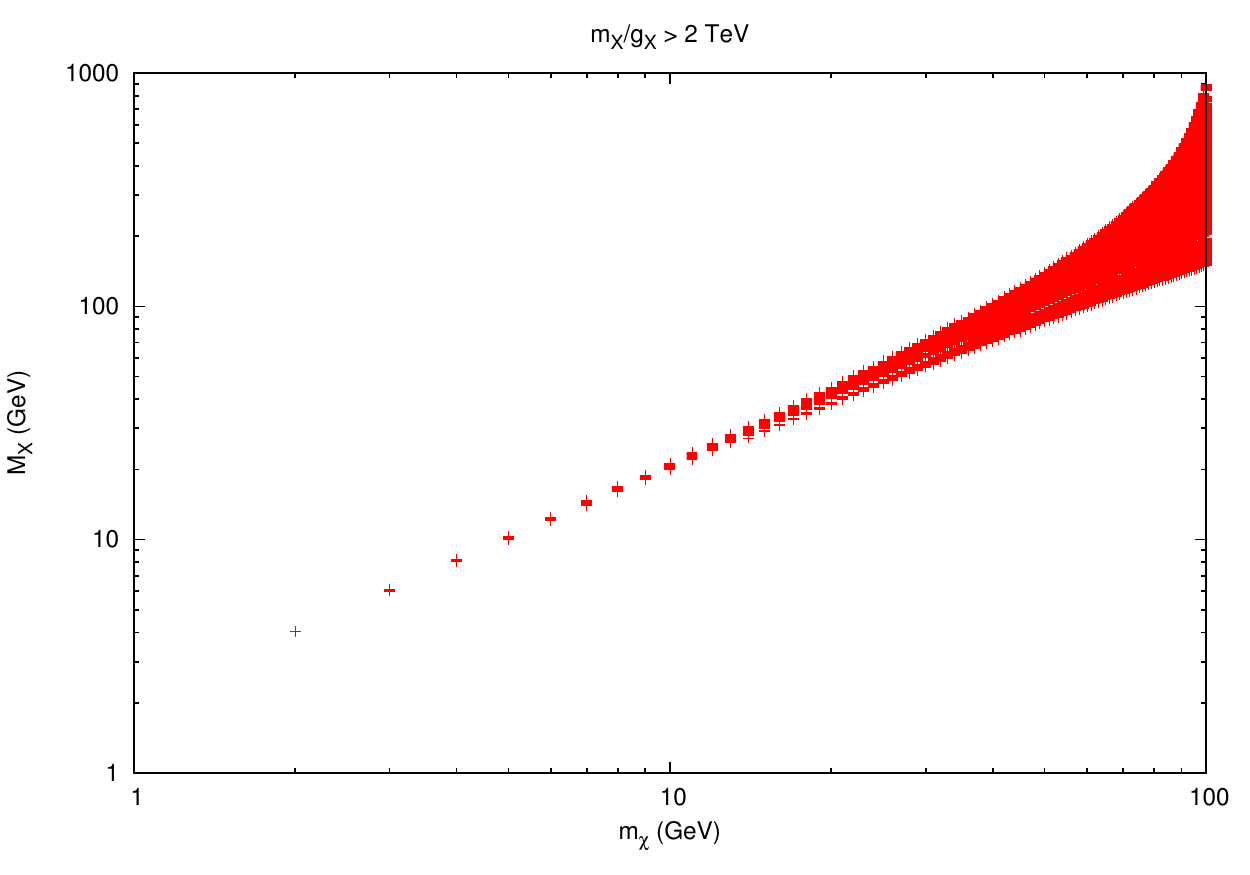} &
\includegraphics[width=0.5\textwidth]{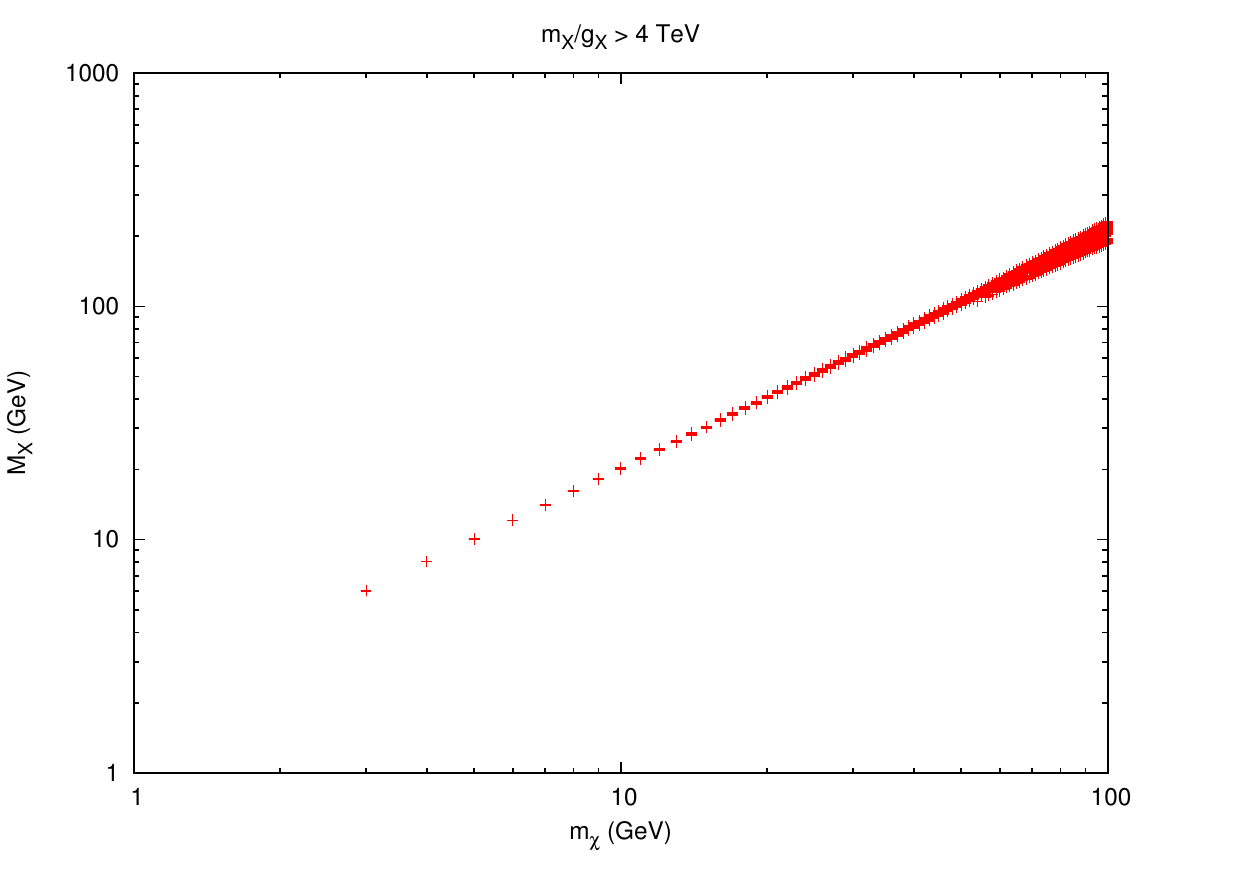}
\end{tabular}
\caption{Fermion Singlet Dark Matter: Points in the $M_X-m_{\chi}$ plane which satisfy the Planck 2013 dark matter relic density bound $\Omega h^2 \in(0.1170,0.1204)$, the constraints on dark matter-nucleon spin dependent cross section from Xenon100 experiment and also the zero-mixing condition (\ref{zeromixeq}). The left and the right plots correspond to $M_X/g_X > 2$ TeV and 4 TeV respectively.}
\label{fig1}
\end{figure}

The analysis of dark matter in this model is similar to that done in \cite{Borah:2012qr} apart from the fact that there is one additional scalar singlet candidate $\chi_2$ for dark matter in this present model. Here we summarise the earlier analysis taking into account of the latest constraints on dark matter relic density from Planck 2013 data \cite{planck}. We also use the latest constraints on spin independent direct detection cross section from LUX 2013 data \cite{LUX} and spin dependent direct detection cross section from Xenon100 experiment \cite{Aprile:2013doa}. In our model there are three fermionic and three scalar candidates for dark matter as they are odd under the remnant $Z_2$ symmetry. Here we briefly discuss the dark matter phenomenology for the singlet fermion and scalar doublet dark matter in the low mass range $10-100$ GeV. The analysis of fermion triplet and scalar singlet dark matter are similar and will be skipped here.

The relic abundance of a dark matter particle $\chi$ is given by the the Boltzmann equation
\begin{equation}
\frac{dn_{\chi}}{dt}+3Hn_{\chi} = -\langle \sigma v \rangle (n^2_{\chi} -(n^{eqb}_{\chi})^2)
\end{equation}
where $n_{\chi}$ is the number density of the dark matter particle $\chi$ and $n^{eqb}_{\chi}$ is the number density when $\chi$ was in thermal equilibrium. $H$ is the Hubble rate and $ \langle \sigma v \rangle $ is the thermally averaged annihilation cross section of the dark matter particle $\chi$. In terms of partial wave expansion $ \langle \sigma v \rangle = a +b v^2$, one obtains the numerical solution of the Boltzmann equation above as \cite{Kolb:1990vq}
\begin{equation}
\Omega_{\chi} h^2 \approx \frac{1.04 \times 10^9 x_F}{M_{Pl} \sqrt{g_*} (a+3b/x_F)}
\end{equation}
where $x_F = m_{\chi}/T_F$, $T_F$ is the freeze-out temperature, $g_*$ is the number of relativistic degrees of freedom at the time of freeze-out. Dark matter particles with electroweak scale mass and couplings freeze out at temperatures approximately in the range $x_F \approx 20-30$. Further simplifications to the solution has been made \cite{Jungman:1995df}:
\begin{equation}
\Omega_{\chi} h^2 \approx \frac{3 \times 10^{-27} cm^3 s^{-1}}{\langle \sigma v \rangle}
\end{equation}
The thermal averaged annihilation cross section $\langle \sigma v \rangle$ is given by \cite{Gondolo:1990dk}
\begin{equation}
\langle \sigma v \rangle = \frac{1}{8m^4T K^2_2(m/T)} \int^{\infty}_{4m^2}\sigma (s-4m^2)\surd{s}K_1(\surd{s}/T) ds
\end{equation}
where $K_i$'s are modified Bessel functions of order $i$, $m$ is the mass of Dark Matter particle and $T$ is the temperature.

\begin{figure}[t]
\centering
\includegraphics[scale=1.0]{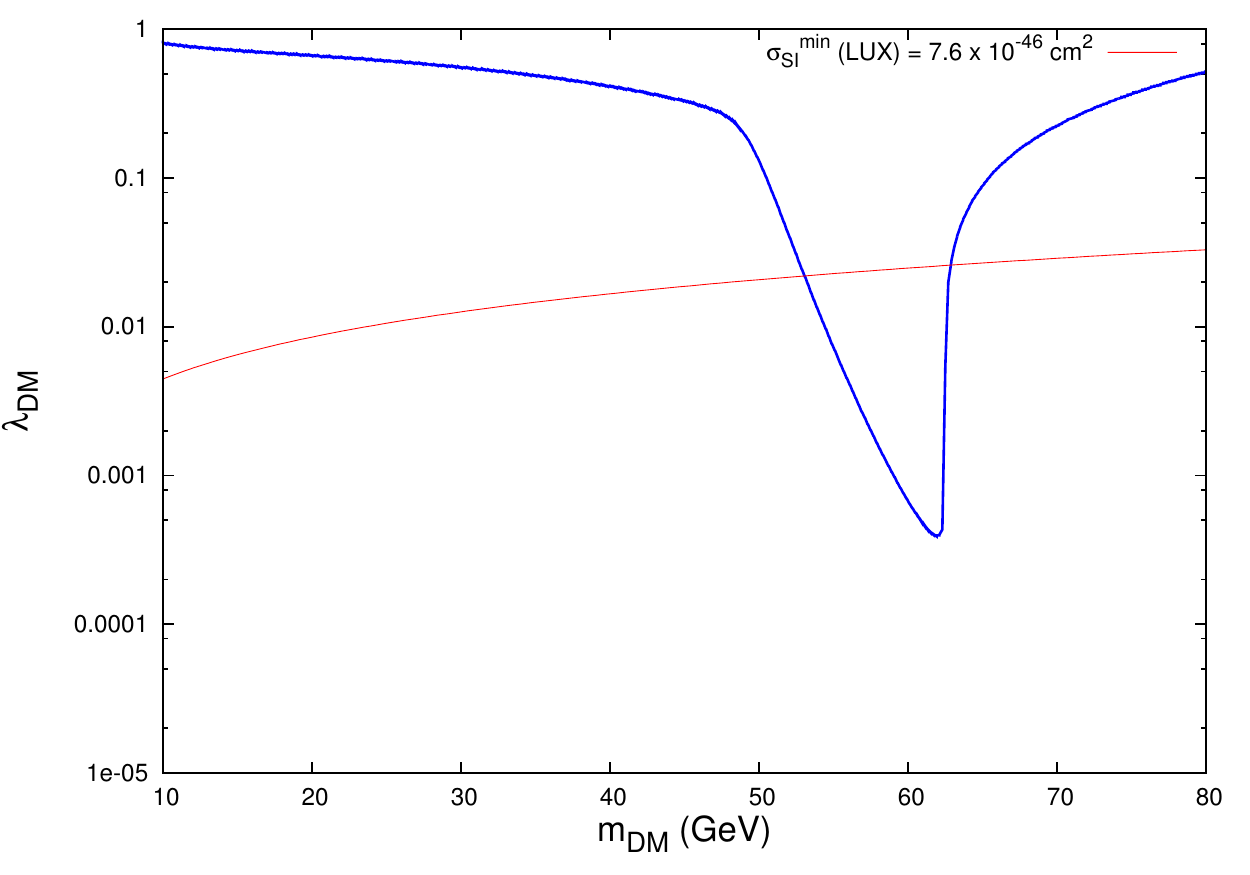}
\caption{Scalar Doublet Dark Matter: Parameter space giving rise to correct Planck 2013 relic density is shown in the $\lambda_{DM}-m_{DM}$ plane. The red line refers to the conservative LUX 2013 upper bound on dark matter nucleon spin independent cross-section.}
\label{fig4}
\end{figure}

There are singlet and triplet  types of fermionic dark matter candidates in our model. As discussed above to calculate relic abundance of any dark matter candidate one is required to identify all possible annihilation processes and find the thermal averaged cross sections. In our model the singlet fermion $N_R$ has no direct coupling to the standard model particles. However, it is charged under the additional gauge group $U(1)_X$. Considering the mixing between X boson and the electroweak gauge bosons, the dominant annihilation channel is the one in which a pair of $N_R$ can annihilate into standard model fermion pairs through s-channel X boson. However, with almost zero mixing condition
as discussed earlier this is not the dominant channel. The Majorana fermion annihilation with s-channel gauge boson mediation was calculated in \cite{Dreiner:2008tw} which have been adopted here for our calculation.

In the annihilation cross-section there are several free parameters  like gauge charges $n_1, n_4$, gauge coupling $g_X$, symmetry breaking scale of $U(1)_X$ as well as the mass of the singlet fermion $N_R$. For the gauge charges, we use the normalization $n^2_1 +n^2_4 = 1$ which was considered earlier by \cite{Adhikari:2008uc,Borah:2012qr} to compute the  exclusion plot  on $M_X/g_X$ at $90\%$ confidence level. In the exclusion plot on $M_X/g_X$  shown in \cite{Adhikari:2008uc} the lowest allowed value of $M_X/g_X$ was found to be approximately $2$ TeV for $\phi = \tan^{-1} (n_4/n_1) = 1.5$. For fixed values of dark matter mass as well as $n_{1,4}$, we vary $g_X$, $u$ and mass of dark matter $m_{\chi}$ and compute the dark matter relic density. The points in the plot \ref{fig1} satisfy the dark matter relic density bound $\Omega_{\chi} h^2 \in (0.1170, 0.1204)$ from Planck experiment \cite{planck} as well as the dark matter direct detection bounds. Being a Majorana fermion, $N_R$ can contribute only to spin dependent dark matter-nucleon cross section and hence one can avoid the stringent bounds on spin-dependent cross section from Xenon100 experiment \cite{Aprile:2013doa}. We follow the analysis of reference \cite{Jungman:1995df} to find out the spin dependent dark matter-nucleon scattering cross section and use the numerical values of nucleon spin fraction carried by the quarks as discussed in reference \cite{pdg}. Xenon100 experiment gives the lowest upper bound on spin dependent cross section as $3.5 \times 10^{-40} \; \text{cm}^2$ for a WIMP mass of 45 GeV at $90\%$ confidence level. Here we take this conservative upper bound for all dark matter mass between 10-100 GeV. 

Similar to our discussion on the fermionic dark matter scenario, the lightest $Z_2$-odd scalar could also be a possible dark matter candidate. Here we discuss in brief the results for the scalar doublet case. This  analysis  is similar to the inert doublet model of dark matter discussed earlier in the literature \cite{Barbieri:2006dq,Cirelli:2005uq,LopezHonorez:2006gr,borahcline, arnabborah14}. We consider the small mass window for the scalar doublet dark matter for which $m_{DM} \leq M_W$, the W boson mass. Beyond the W boson mass threshold, the annihilation channel of scalar doublet dark matter into $W^+W^-$ pairs opens up resulting in the reduction of  the relic abundance of dark matter below observed range for dark matter mass approximately below $500$ GeV. One may note however, that there exists a region of parameter space $M_W < m_{DM} <160$ GeV which satisfy relic density bound if certain cancellations occur between several annihilation diagrams \cite{honorez1}. For the sake of simplicity, we stick to particularly the low mass region $10 \; \text{GeV} < m_{DM} < M_W$ in this analysis. The annihilation cross section of scalar doublet dark matter $\phi_3$ into fermion-antifermion pairs is computed through standard model like Higgs boson $(\text{mass} m_h = 126 \; \text{GeV})$. As the standard model Higgs boson coupling to fermions are well known, 
essentially the only free parameters in the annihilation cross section are the dark matter-Higgs coupling $\lambda_{DM}$ and dark matter mass $m_{DM}$. The v-shaped blue region in figure \ref{fig4} shows the parameter space in $\lambda_{DM}-M_{DM}$ plane which satisfies the relic density bound. To implement the direct detection constraints on spin independent scatterring cross section from LUX experiment \cite{LUX}, we use the Higgs boson mediated spin independent scattering cross section \cite{Barbieri:2006dq}
\begin{equation}
 \sigma_{SI} = \frac{\lambda^2_{DM}f^2}{4\pi}\frac{\mu^2 m^2_n}{m^4_h m^2_{DM}}
\label{sigma_dd}
\end{equation}
where $\mu = m_n m_{DM}/(m_n+m_{DM})$ is the DM-nucleon reduced mass. We use the recent estimate of the Higgs-nucleon coupling $f = 0.32$ \cite{Giedt:2009mr} although the full range of allowed values is $f=0.26-0.63$ \cite{mambrini}. The red solid line in figure \ref{fig4} corresponds to the minimum upper limit on the dark matter-nucleon spin independent cross section of $7.6 \times 10^{-46} \; \text{cm}^2$ from LUX experiment \cite{LUX}.
\section{Results and Conclusion}
\label{sec:results}
Motivated by the tantalizing hints from cosmology as well as neutrino oscillation experiments favouring the possible existence 
of sterile neutrinos, we have studied an abelian extension of the standard model with anomaly free combination of three right handed singlet and two triplet fermions. The abelian gauge symmetry gets spontaneously broken down to a remnant $Z_2$ symmetry such that the lightest $Z_2$ odd particle could play the role of dark matter. The scalar sector of the model is chosen in such a way that three active and one sterile neutrinos can acquire masses at electron Volt scale. At tree level, only one active neutrino acquires masses through type I seesaw mechanism whereas two active and one sterile neutrino remain massless. At one loop level, all the three active and one sterile neutrinos receive non-zero mass contribution at eV scale. Due to the loop suppression factor, the new physics scale could be as low as TeV scale without paying the price of fine-tuning the dimensionless parameters. This could have interesting signatures in the collider experiments. This model also allows non-zero mixing between active and sterile neutrinos at loop level and hence can provide a suitable explanation to reactor neutrino anomalies. 

Due to the remnant $Z_2$ symmetry in our model, the lightest $Z_2$ odd particle is stable and hence can play the role of dark matter. We briefly discuss the case of fermion singlet and scalar doublet dark matter in the low mass range $10-100$ GeV and show how the latest experimental data can be satisfied within the framework of our model. Although the original motivation of our work is to put forward a common mechanism for masses and mixing of eV scale active and sterile neutrino  with dark matter, this model can also be studied in order to explain keV scale sterile neutrino which can play the role of warm dark matter. Although the light sterile neutrino in our model is not protected by the remnant $Z_2$ symmetry, its mass and mixing with active neutrinos can be suitably adjusted in such a way that its lifetime exceeds the age of the Universe. This will give rise to a mixed dark matter scenario where the lightest $Z_2$ odd particle acts as a cold dark matter candidate whereas the keV scale sterile neutrino acts as a warm dark matter candidate. We leave a detailed study of dark matter phenomenology in this scenario to future studies.

\section{Acknowledgement}

RA thanks Department of Physics, UC, Riverside for the kind hospitality during the initial stage of this work.

\bibliographystyle{apsrev}
%\bibliography{ref_abelian_dm,scalarDM,seesaw,ref4thGen}

\end{document}